\input harvmac
\input epsf

\Title{
\rightline{hep-th/0205097}}
{\vbox{\centerline{Non-maximally symmetric D-branes 
on group manifold}
\centerline{ in the Lagrangian approach}}}

\medskip
\centerline{Gor Sarkissian\foot{e-mail: gor@ictp.trieste.it}}
\bigskip
\smallskip
\centerline{The Abdus Salam International Centre for Theoretical Physics}
\centerline{Strada Costiera 11, Trieste 34014, Italy}

\smallskip

\bigskip\bigskip\bigskip
\noindent
Recently, Maldacena, Moore and Seiberg introduced non-maximally symmetric 
boundary states on group manifold using T-duality. In the work 
presented here we
suggest simple description of these branes in terms of group elements.
We show that T-dualization actually reduces to multiplication
of conjugacy classes by the corresponding $U(1)$ subgroups. 
Using this description we find the two-form trivializing the WZW three-form on
 the branes. $SU(2)$ and $SL(2,R)$ examples are considered in details.

\vfill
\Date{5/02}

\newsec{Introduction}
\lref\gabrw{M. R. Gaberdiel, A. Recknagel and G. M. T. Watts,
The conformal boundary states for $SU(2)$ at level $1$,
Nucl.Phys. B626 (2002) 344, hep-th/0108102.}

\lref\gabreck{M. R. Gaberdiel and A. Recknagel,
Conformal boundary states for free bosons and fermions,
JHEP 0111 (2001) 016, hep-th/0108238.}

\lref\jan{R. A. Janik, Exceptional boundary states at $c=1$, 
Nucl.Phys. B618 (2001) 675-688, 
hep-th/0109021.} 

\lref\capap{A. Cappelli and G. D'Appollonio, 
Boundary states of $c=1$ and $c=3/2$ rational conformal field theories,
JHEP 02 (2002) 039, hep-th/0201173.}

\lref\klsev{C. Klimcik and P. Severa, 
Open strings and D-branes in WZNW models,
Nucl. Phys. B488 (1997) 653, hep-th/9609112.}%

\lref\alschom{A. Alekseev and V. Schomerus, 
D-branes in the WZW model, Phys. Rev. D60, 061901 (1999),
hep-th/9812193.}%

\lref\gaw{K. Gawedzki, 
Conformal field theory: a case study, in Conformal Field Theory,
Frontier in Physics 102, Eds. Y. Nutku, C. Saclioglu, T. Turgut
( Perseus Publishing, 2000),
hep-th/9904145.}%

\lref\stanold{S. Stanciu, 
A note on D-branes in group manifolds: flux quantization and D0-charge,
JHEP 0010 (2000) 015, hep-th/0006145.}%

\lref\bfs{L. Birke, J. Fuchs and C. Schweigert, hep-th/9905038.}%

\lref\gcpleb{H. Garcia-Compean and J.F. Plebanski, hep-th/9907183.}%

\lref\arsold{A. Alekseev, A. Recknagel and V. Schomerus, hep-th/9908040,
JHEP {\bf 9909} (1999) 023.}%

\lref\fffs{G. Felder, J. Frohlich, J. Fuchs and C. Schweigert,
hep-th/9909030.}%
\lref\fofs{J.M. Figueroa-O'Farrill and S. Stanciu, hep-th/0001199.}%
\lref\bdsnew{C. Bachas, M. Douglas and C. Schweigert, 
Flux stabilization of D-branes, JHEP 0005 (2000) 048,
hep-th/0003037.}%
\lref\arsnew{A. Alekseev, A. Recknagel and V. Schomerus, hep-th/0003187.}%
\lref\fredshom{S.\ Fredenhagen and V.\ Schomerus, hep-th/0012164.}%
\lref\malmnat{J.\ Maldacena, G.\ Moore and N.\ Seiberg,
D-brane instantons and K-theory charges, JHEP 0110 (2001) 005,
 hep-th/0108100.}%
\lref\malmosei{J. Maldacena, G. Moore and N. Seiberg, 
Geometrical interpretation of D-branes in gauged WZW models,
JHEP 0107 (2001) 046,
hep-th/0105038.}%
\lref\cardy{J.L. Cardy, Nucl. Phys. B324 (1989) 581.}%
\lref\gawtod{K. Gawedzki, I. Todorov, P. Tran-Ngoc-Bich,
Canonical quantization of the boundary Wess-Zumino-Witten model,
hep-th/0101170.}
 \lref\ngaw{K. Gawedzki, 
Boundary WZW, G/H, G/G and CS theories,
hep-th/0108044.}%
\lref\elsar{S. Elitzur and G. Sarkissian, 
D-branes on a gauged WZW model,
Nucl.Phys. B625 (2002) 166-178,
hep-th/0108142.}
\lref\egkrs{S. Elitzur, E. Rabinovici, A. Giveon, D. Kutasov
and G. Sarkissian, 
D-branes in the Background of NS Fivebranes,
JHEP 0008 (2000) 046, hep-th/0005052.}
\lref\quschom{T. Quella and V. Schomerus, 
Symmetry breaking boundary states and defect lines,
hep-th/0203161.}
\lref\rajar{A. Rajaraman, 
New $AdS_3$ Branes and Boundary States,
hep-th/0109200.}
During the last years, D-branes were studied using 
boundary conformal field theory.
One of the most important criteria used in these studies
was the number of symmetries preserved by the D-branes in question.
Mainly were studied branes preserving the full chiral
symmetry, which is usually broader than conformal symmetry.
 
Recently, some attempts were made to study branes preserving
smaller symmetries. In \gabrw,\ \gabreck,\ \jan,\ \capap\  were constructed
boundary states preserving only conformal symmetry.

In \malmosei,\ \malmnat,\  was developed a general approach
for constructing boundary states of WZW models preserving only some part
of the full affine symmetry by use of the T-duality
in the directions of the Cartan subalgebra.
Further, for the non-abelian subgroups, this approach 
was developed in \quschom\ .
But whenever a CFT target space receives a geometrical 
interpretation, the algebraically constructed brane can
be realized as a geometrical subset. 
In \alschom\ , maximally symmetric branes in the WZW model
are given by finite number of conjugacy classes.
In \malmosei\ the shape
of the non-maximally symmetric branes for the case of SU(2) group manifold
was found, but
this description left obscured the connection to underlying symmetries 
as well as, the possibility of generalization to other groups.
An attempt to find similar  branes on the $SL(2,R)$
was done in \rajar\ .

In this work we suggest simple description of the non-maximally symmetric
$D$-branes, derived by means of T-duality,
in terms of group elements using symmetry arguments.
We show that they are given by product of conjugacy classes with T-dualized 
$U(1)$ subgroups.

We show consistent with work performed in  \klsev , \alschom\ and \gaw\ ,
that, in these branes, the WZW three-form belongs to the trivial 
cohomology class. We construct action with this boundary condition
and show that it displays required symmetry.

We show that this action also can be derived by the direct
T-dualization of the boundary WZW action.

Finally we show that for $G=SU(2)$ location and mass of these branes
coincide with corresponding values found in \malmosei\ .
We also analyzed in detail, the 
location of these branes for the case of SL(2,R).
\newsec{Algebraic description of T-dualized branes}
In this section we briefly review construction
 of the non-maximally symmetric boundary state, introduced
in \malmosei\ and \malmnat\ .

The main idea of the construction is to represent group manifold $G$
as orbifold $G=(G/H\times H)/\Gamma$, where $H$ is an abelian subgroup of $G$, 
and then constructing
boundary state for $G$ as $\Gamma$-invariant 
linear combination of the tensors product of boundary states
for coset $G/H$ and for abelian subgroup $H$. 

 Let us illustrate this
abstract construction for the case, when $G=SU(2)_k$,
 $H=U(1)_k$, $G/H=PF_k$ and $\Gamma=Z_k$: 
\eqn\pardec{SU(2)_k=(PF_k\times U(1)_k)/Z_k.}

Before describing the results in depth, it may prove useful to review
briefly the boundary states for $U(1)_k$.
Here as usual for the scalar field, we have
the Neumann and Dirichlet boundary states. However
extended symmetry present at the special values of the radius 
imposes some restrictions on the position of $D0$-brane for the
Dirichlet boundary condition and on the Wilson line parameter 
of the $D1$-brane for the Neumann boundary condition.
Using accepted notation $A$-branes and $B$-branes for the Dirichlet
and Neumann boundary conditions correspondingly, the boundary states
found in \malmosei\ are:
\eqn\abrane{|A,n\rangle={1\over (2k)^{1/4}}\sum_{n'=0}^{2k-1}
e^{{-i\pi nn'\over k}}|A,n',n'\rangle\rangle,}    
for A-branes, and
\eqn\bbrane{|B,\eta=\pm 1\rangle=({k\over 2})^{1/4}[|B,0,0\rangle\rangle
+\eta|B,k,-k\rangle\rangle],}
for B-branes, where $|A,n',n'\rangle\rangle$ and $|B,r,-r\rangle\rangle$
are $A$- and $B$-Ishibashi states correspondingly:
 
\eqn\aish{|Ar,r\rangle\rangle=\exp\left[
+\sum_{n=1}^{\infty}{\alpha_{-n}\tilde{\alpha}_{-n}\over n}\right]
\sum_{l\in Z}|{r+2kl\over \sqrt{2k}},{r+2kl\over \sqrt{2k}}
\rangle,}

\eqn\bish{|Br,-r\rangle\rangle=\exp\left[
-\sum_{n=1}^{\infty}{\alpha_{-n}\tilde{\alpha}_{-n}\over n}\right]
\sum_{l\in Z}|{r+2kl\over \sqrt{2k}},-{r+2kl\over \sqrt{2k}}
\rangle.}
We see that \abrane\ describes $D0$-brane sitting at $2k$ special points,
 and \bbrane\ are two $D1$-branes with special
values of the Wilson line parametrized by $\eta$.

Now we are in position to write down two kinds of boundary states for the
$SU(2)_k$ according to the main prescription:
\eqn\asubr{|A,j\rangle={1\over \sqrt{k}}\sum_n|Ajn\rangle^{{\rm PF}}
|A,n\rangle}
and 
\eqn\bsubr{|B,j,\eta=\pm 1\rangle={1\over \sqrt{k}}
|B,\eta\rangle\sum_{n=0}^{2k-1}|Ajn\rangle^{{\rm PF}},}
where $|Ajn\rangle^{{\rm PF}}$ is the usual Cardy state 
for the parafermionic theory.
It is easy to check that the Dirichlet  gluing condition
\asubr\ just gives us the usual maximally symmetric Cardy state 
for the $SU(2)_k$ affine algebra, but \bsubr\ gives us new 
non-maximally symmetric branes. From the form of the boundary state
we see that preserved symmetry now is diagonal $U(1)$ (and a $Z_k$).
 Using formula \bsubr\ it is easy to compute mass and shape
of the new branes. 

In \malmosei\ the mass was found by the overlap
of the \bsubr\  with $|A,j=0\rangle\rangle$:
\eqn\mass{M(Bj)=\sqrt{k}M(Aj).}

Then by the overlap of the \bsubr\ with the graviton wave packet
was found the shape of the branes:
\eqn\loc{\langle B,j,\eta|\tilde{\theta}\rangle\sim
k\sum_jD_{00}^{j'}S_{jj'}\sim ie^{i\hat{\psi}}\sum_{n=0}^{\infty}
P_n(\cos\tilde{\theta})e^{in2\hat{\psi}}+{\rm c.c.}
\sim {\Theta(\cos\tilde{\theta}-\cos2\hat{\psi})\over
\sqrt{\cos\tilde{\theta}-\cos2\hat{\psi}}},}
where $D_{mm'}^{j'}$ is matrix of rotations, $P_n$ are Legendre polynomials,
$\tilde{\theta}$ is the second Euler angle, $\hat{\psi}={2j\pi\over k}$,
$S_{jj'}$ is matrix of modular transformation for $SU(2)_k$,
and $\Theta(z)$ is the usual step function which vanishes when
$z<0$. 
We see that generically these are three-dimensional branes covering only
 part of the group manifold. But for even values of $k$, T-dualizing
the biggest equatorial conjugacy class with $j={k\over 4}$ results in brane 
covering the whole group manifold. In \malmosei\ it was 
conjectured that the partially covering branes are unstable, 
while the last one is stable.
By T-dualizing D0-brane, a D1-brane is formed along a maximum
circle of $S^3$, which is unstable.

It is straightforward to generalize construction of the \bsubr\ 
to that of other groups \malmnat\ .
Let us consider for example unitary group $G=SU(N+1)$ and
the embedding $SU(N)\times U(1)\hookrightarrow SU(N+1)$, 
where $U(1)$ corresponds to
the generator $H^{N}={1\over \sqrt{N(N+1)}}{\rm Diag}\{1,\dots,1,-N\}$.
 It was shown in \malmnat\ 
that performing $T$-duality with respect
to the current $H^{N}$ we get boundary state $|B\rangle$,
 satisfying the boundary conditions:
\eqn\bouncond{
(J^a_n+\tilde{J}^a_{-n})|B\rangle=0\;\; a\in su(N),}
\eqn\boundu{ (H^N_n-\tilde{H}^N_{-n})|B\rangle=0.}
However this  does not satisfy simple boundary conditions with respect to 
the remaining currents.
Therefore preserved symmetry is now 
$SU(N)_{\rm vectorial}\times U(1)_{\rm axial}$, where subscripts vectorial
and axial refer to the signs plus and minus in \bouncond\ and \boundu\
respectively.

\newsec{Geometrical description of the non-maximally symmetric branes}
\subsec{Definition} 
In this section we present the main result of this work, which is 
the geometrical
description of the T-dualized boundary state reviewed in the
previous section. 
 
It is useful to begin with the Polyakov-Wiegmann
identities which will be referred to frequently in this section.
\eqn\pwk{ L^{\rm kin}(g h)=  L^{\rm kin}(g) + L^{\rm kin} (h)
 -( {\rm Tr} (g^{-1}\partial_z g \partial_{\bar z} h h^{-1}+
 {\rm Tr} (g^{-1} \partial_{\bar z} g\partial_z  h h^{-1})),}

\eqn\pwwz{\omega^{\rm WZ}(g h) = \omega^{\rm WZ}(g) + \omega^{\rm WZ}(h)
 - d({\rm Tr} (g^{-1} dg  dh h^{-1})),}

where $L^{\rm kin}={\rm Tr}(\partial_zg\partial_{\bar{z}}g^{-1})$
and $\omega^{\rm WZ}={1\over 3}{\rm Tr}(g^{-1}dg)^3$. 

We define the new D-branes as product of the conjugacy class
with the U(1) subgroup. In other words
we study  the following boundary condition:
\eqn\newbond{g_{|{\rm boundary}}=Lhfh^{-1},}
 where $L \in U(1)$, $f=e^{{2\pi\over
k}\lambda\cdot H}$,  $H$  are Cartan generators, $\lambda$ is a vector in the
 weight lattice, $h \in G$.
We denote $C=hfh^{-1}$.

Recalling that, according to the analysis of \klsev\ , and
\gaw\ , in order for some subset to be a good boundary condition
for the WZW model, restriction of the WZW three-form to that subset
should belong to the trivial cohomology class. In other words,
on the subset should exist a two-form $\omega^{(2)}$ satisfying the 
condition:
\eqn\trcohom{d\omega^{(2)}=\omega^{{\rm WZ}}.}
It may be easily checked that proposed branes satisfy those criteria.
At the boundary \newbond\ according to \pwwz\
\eqn\wzontd{\omega^{\rm WZ}(g)=\omega^{\rm WZ}(L)+\omega^{\rm WZ}(C)
-d{\rm Tr}(L^{-1}dLdCC^{-1}).}
Using that for the abelian group,  $L$,  $\omega^{\rm WZ}(L)=0$,
  and  
\eqn\wzformm{\omega^{\rm WZ}(C)=d\omega^f(h)=
d{\rm Tr}(h^{-1}dhfh^{-1}dhf^{-1}),}
we get 
\eqn\wzontdd{\omega^{\rm WZ}(g)_{|{\rm boundary}}=d\omega^{(2)}(L,h),}
where
\eqn\wzonbound{
\omega^{(2)}(L,h)=\omega^f(h)-{\rm Tr}(L^{-1}dLdCC^{-1}).}

\subsec{Action and symmetry check}
The action is defined as:
\eqn\moactn{ S ={ k_{G}\over{4 \pi}}\left[\int_{\Sigma}d^2z L^{\rm kin} 
+ \int_B \omega^{\rm WZ}-
\int_{D}\omega^{(2)}(L,h)\right],}
where $\partial B=\Sigma+D$.

The boundary is invariant under the following list of
transformations.
\item{(1)}
$g\rightarrow kgk^{-1}$ satisfying to the condition $[k,L]=0$.
Under this transformation $h\rightarrow kh$ and $C\rightarrow kCk^{-1}$.
 This means that for example, in  
the case of $SU(N+1)$, $k\in SU(N)\times U(1)$.

\item{(2)}
$g\rightarrow kg$, where $k\in U(1)$. Under this transformation
$L\rightarrow kL$. 

\item{(3)}
$g\rightarrow gk$, where $k\in U(1)$. Under this transformation
$L\rightarrow Lk$, $C\rightarrow k^{-1}Ck$ and $h\rightarrow k^{-1}h$.
It follows from (2) and (3)  that the boundary is invariant
also under their axially diagonal combination:
\item{(4)}
$g\rightarrow kgk$, where $k\in U(1)$. Under this transformation
$L\rightarrow kLk$, $C\rightarrow k^{-1}Ck$ and $h\rightarrow k^{-1}h$.
  
Using the method developed in \elsar\ , now we will show, 
that the action \moactn\ is  invariant under the following symmetries:
\item{(1)}
\eqn\sym{g(z,\bar{z})\rightarrow  h_L(z) g(z,\bar{z}) h_R^{-1}(\bar {z}),}
with $h_L(z)|_{\rm boundary}=h_R(\bar {z})|_{\rm boundary}=k(\tau)$ 
, $k\in SU(N)$, in agreement with \bouncond\ .
\item{(2)}
\eqn\symu{g(z,\bar{z})\rightarrow  h_L(z) g(z,\bar{z}) h_R(\bar {z}),}
with $h_L(z)|_{\rm boundary}=h_R(\bar {z})|_{\rm boundary}=k(\tau)$,
$k\in U(1)$, in agreement with \boundu\ . 
It is important to note that in \sym\ we used vectorial combination of
the left and right symmetries, whereas in \symu\ axial combination
is used, in agreement with the sign difference between \bouncond\ 
and \boundu\ .

Under the transformation \sym, the change
in the $L^{\rm kin}$ term in  \moactn\ read from \pwk\ ,  is canceled
 by the corresponding $\Sigma$ integral of the  boundary term
 from the change in the $ \omega^{\rm WZ}$ term, read from \pwwz\ .
 In the presence
 of a world sheet  boundary there remains the contribution from $D$ to 
the latter change, 

\eqn\delwzal{\Delta (S^{\rm kin}+S^{\rm WZ}) = 
{ k_G\over{4 \pi}} \int_D {\rm Tr}[k^{-1}dk( g k^{-1}dk g^{-1} 
- g^{-1}dg - dg g^{-1})],}
where $g=LC$.
Substituting this value of $g$ in \delwzal\ we get: 
\eqn\delwzall{\eqalign{&\Delta (S^{\rm kin}+S^{\rm WZ}) = 
{ k_G\over{4 \pi}} \int_D {\rm Tr}[k^{-1}dk( LCk^{-1}dk C^{-1}L^{-1}
- C^{-1}L^{-1}(dLC+LdC)-\cr
&(dLC+LdC)C^{-1}L^{-1})],}}
and using $[k,L]=0$ and cyclic permutation under the trace we obtain:
\eqn\delwzalbr{\Delta (S^{\rm kin}+S^{\rm WZ}) = 
{ k_G\over{4 \pi}} \int_D {\rm Tr}[k^{-1}dk(Ck^{-1}dk C^{-1}
-C^{-1}L^{-1}dLC-C^{-1}dC -dLL^{-1}-dCC^{-1})].}
Now we compute $\omega^{(2)}(L,kh) -\omega^{(2)}(L,h)$,
using that 
\eqn\omegtran{\omega^f(kh)-\omega^f(h)={\rm Tr}[k^{-1}dk(Ck^{-1}dkC^{-1}
-C^{-1}dC-dCC^{-1})]}
and
\eqn\secpart{{\rm Tr}[L^{-1}dLd(kCk^{-1})kC^{-1}k^{-1}-
L^{-1}dLdCC^{-1}]={\rm Tr}[L^{-1}dLdkk^{-1}-L^{-1}dLCk^{-1}dkC^{-1}],}
resulting in
\eqn\trfin{\eqalign{&\omega^{(2)}(L,kh) -\omega^{(2)}(L,h)
={\rm Tr}[k^{-1}dk(Ck^{-1}dkC^{-1}
-C^{-1}dC-dCC^{-1}+L^{-1}dL-\cr
&C^{-1}L^{-1}dLC)].}}
Collecting \delwzalbr\ and \trfin\ we obtain:
\eqn\vecact{\Delta S={ k_G\over{2 \pi}} \int_D {\rm Tr}(L^{-1}dLk^{-1}dk).}
Noting, that for $k\in SU(N)$ and $L\in U(1)$
${\rm Tr}(L^{-1}dLk^{-1}dk)=0$, we prove that the action \moactn\ possesses
by the vectorially diagonal $SU(N)$ symmetry. We also see from  \vecact\
that the vectorially diagonal $U(1)$ symmetry is broken.

Now we will show that the action \moactn\ possesses by 
the axially diagonal $U(1)$
symmetry \symu\ .
By the same arguments, leading to the \delwzal\ , we get that
in the presence of the boundary under \symu\ :
\eqn\delwzaluu{\Delta (S^{\rm kin}+S^{\rm WZ}) = 
{ k_G\over{4 \pi}} \int_D {\rm Tr}[k^{-1}dk( g^{-1}dg-g k^{-1}dk g^{-1} 
- dg g^{-1})],}
where $g=LC$.
Substituting this value of $g$ in \delwzaluu\ we get: 
\eqn\delwzallu{\Delta (S^{\rm kin}+S^{\rm WZ}) = 
{ k_G\over{4 \pi}} \int_D {\rm Tr}[k^{-1}dk
(C^{-1}dC-Ck^{-1}dk C^{-1}
+C^{-1}L^{-1}dLC-dLL^{-1}-dCC^{-1})].}
Now we compute $\omega^{(2)}(kLk,k^{-1}h) -\omega^{(2)}(L,h)$,
using that 
\eqn\omegtranu{\omega^f(k^{-1}h)-\omega^f(h)={\rm Tr}[k^{-1}dk(Ck^{-1}dkC^{-1}
+C^{-1}dC+dCC^{-1})]}
and
\eqn\secpartu{\eqalign{
&{\rm Tr}[(kLk)^{-1}d(kLk)d(k^{-1}Ck)k^{-1}C^{-1}k-
L^{-1}dLdCC^{-1}]=\cr
&{\rm Tr}[k^{-1}dk(2dCC^{-1}+2Ck^{-1}dkC^{-1}+L^{-1}dL-C^{-1}L^{-1}dLC)],}}
resulting in
\eqn\trfin{\eqalign{&\omega^{(2)}(L,kh) -\omega^{(2)}(L,h)
={\rm Tr}[k^{-1}dk(C^{-1}dC-Ck^{-1}dkC^{-1}
-dCC^{-1}-L^{-1}dL+\cr
&C^{-1}L^{-1}dLC)],}}
which cancels \delwzallu\ .

\subsec{T-duality}

\lref\kirit{E. Kiritsis,
Exact Duality Symmetries in CFT and String Theory,
Nucl.Phys. B405 (1993) 109-142,
 hep-th/9302033.}
In this subsection we give  alternative derivation of the form
$\omega^{(2)}(L,h)$ explaining its relation to the T-duality.

Remembering how to get the T-dual WZW action in the absence of boundary,
 as shown in \kirit\ , 
we parametrise group element as a product 
\eqn\tdpar{g=L^{-1}p=e^{-i\phi H^{N}}p,}
where $H^{N}$ is a generator of the Lie algebra, then, using
the Polyakov-Wiegmann identity, separate $\phi$ and $p$ parts,
and afterwards T-dualise the scalar part.
In the presence of the boundary as we will see this procedure will 
be modified by boundary terms.

Considering  boundary WZW action with conjugacy class as boundary
condition,  $g_{|{\rm boundary}}=hfh^{-1}$:
\eqn\moactbound{ S ={ k_{G}\over{4 \pi}}\left[\int_{\Sigma}d^2z L^{\rm kin} 
+ \int_B \omega^{\rm WZ}-
\int_{D}\omega^{f}(h)\right].}
As was established in \alschom\ and \gaw\ with $f=e^{{2\pi\over
k}\lambda\cdot H}$, where $H$  are Cartan generators, and $\lambda$ is
 a vector in the weight lattice, \moactbound\ is a well defined action.

Inserting g in the form \tdpar\ to \moactbound\
after using the Polyakov-Wiegamnn identities \pwk\ , \pwwz\
we obtain:
\eqn\moactboundtd{\eqalign{& S={k_{G}\over{4 \pi}}
[\int_{\Sigma}d^2z L^{\rm kin}(p) 
+ \int_B \omega^{\rm WZ}(p)-
\int_{D}(\omega^{f}(h)-{\rm Tr}(L^{-1}dLdpp^{-1}))\cr 
&+\int_{\Sigma}\partial_{z}\phi
\partial_{\bar{z}}\phi-2i\int_{\Sigma}
\partial_{\bar{z}}\phi{\rm Tr}(H^{N}\partial_{z}pp^{-1})].}}

At this point, boundary condition for $\phi$
should be specified.
From \tdpar\ we see that if $\phi$ satisfies to the Dirichlet boundary
condition, $p$ at the boundary lies in the usual conjugacy class, but
if $\phi$ satisfies to the Neumann boundary condition, $p$
at the boundary takes value $p=Lhfh^{-1}$. In other words, it lies in the
above discussed branes \newbond\ .

After short algebra it may be checked, 
that the integrand of the boundary integral in 
\moactboundtd\ equals to \wzonbound\ :
\eqn\derforn{\eqalign{
&{\rm Tr}(L^{-1}dLdpp^{-1})={\rm Tr}(L^{-1}dLd(LC)C^{-1}L^{-1})=
{\rm Tr}(L^{-1}dL(dLC+LdC)C^{-1}L^{-1})=\cr
&{\rm Tr}(L^{-1}dLdLL^{-1})+{\rm Tr}(L^{-1}dLdCC^{-1})=
{\rm Tr}(L^{-1}dLdCC^{-1}),}}
where we used that ${\rm Tr}(L^{-1}dLdLL^{-1})=0$ for the abelian group.
This computation shows that  \moactboundtd\ is actually
the sum of the action \moactn\ ,with the new branes as the boundary condition, 
with scalar field coupled to current.
Since, as noted above, the action \moactbound\ , for $f$ chosen as above, is
well-defined WZW action, it is proven that with
the same choice of $f$ also \moactn\ is well defined WZW action.
\lref\vil{ N. J. Vilenkin, Special functions and the theory of group
representations, AMS.}
\newsec{Examples}
\subsec{Branes on SU(2)}
Let us consider now the case $g=SU(2)$ in details.
It is convenient to parametrise the group element as
\eqn\grel{g=x_0\sigma_0+i(x_1\sigma_1+x_2\sigma_2+x_3\sigma_3),}
with $x_0^2+x_1^2+x_2^2+x_3^2=1$.
In this parametrisation conjugacy class given as $x_0=\cos\hat{\psi}$
where $\hat{\psi}={2\pi j\over k}$.
This parametrisation connected with the Euler angles 
\eqn\euang{g=e^{i\chi{\sigma_3\over 2}}e^{i\tilde{\theta}{\sigma_1\over 2}}
e^{i\phi{\sigma_3 \over 2}}}
by formulae
\eqn\eulerang{\eqalign{
& x_0=\cos{\tilde{\theta}\over 2}\cos{\chi+\phi\over 2}\cr
& x_1=\sin{\tilde{\theta}\over 2}\cos{\chi-\phi\over 2}\cr
& x_2=\sin{\tilde{\theta}\over 2}\sin{\chi-\phi\over 2}\cr
& x_3=\cos{\tilde{\theta}\over 2}\sin{\chi+\phi\over 2}.}}

We note that  $x_0^2+x_3^2=\cos^2{\tilde{\theta}\over 2}$
and $x_1^2+x_2^2=\sin^2{\tilde{\theta}\over 2}$.

If one parametrises the $U(1)$ subgroup as $e^{i\alpha\sigma_3}$
the D-branes are located at:
\eqn\brnloc{
e^{i\alpha\sigma_3}(\cos\hat{\psi}\sigma_0+i(x_1\sigma_1+x_2\sigma_2+
x_3\sigma_3))
=\tilde{x}_0\sigma_0+i(\tilde{x}_1\sigma_1+\tilde{x}_2\sigma_2+
\tilde{x}_3\sigma_3),}
where
\eqn\newcor{\eqalign{
&\tilde{x}_0=\cos\hat{\psi}\cos\alpha-x_3\sin\alpha \cr
&\tilde{x}_1=x_1\cos\alpha+x_2\sin\alpha \cr
&\tilde{x}_2=x_2\cos\alpha-x_1\sin\alpha \cr
&\tilde{x}_3=x_3\cos\alpha+\cos\hat{\psi}\sin\alpha.}}

We see that $\tilde{x}^2_1+\tilde{x}_2^2=x_1^2+x_2^2$.
From one side as we noted above $\tilde{x}_1^2+\tilde{x}_2^2=
\sin^2{\tilde{\theta}\over 2}$,
from the other side maximum value of the $x_1^2+x_2^2$ on the 
conjugacy class is obviously $\sin^2\hat{\psi}$.
So we have that on the new branes 
\eqn\brcond{\sin^2{\tilde{\theta}\over 2}\leq\sin^2\hat{\psi}.}
Using that $2\sin^2\alpha=1-\cos 2\alpha$,
we get that on the branes
\eqn\brlocf{\cos\tilde{\theta}\geq\cos 2\hat{\psi},}
which is exactly \loc.
It is useful to think about the new D-branes also as a 
collection of translated conjugacy classes along the whole
$U(1)$ subgroup. From this we get that their volume
equals to the product of the radius of the $U(1)$
subgroup and the volume of the conjugacy class. 
This perfectly matches to the mass formula \mass\ .
We also note that for $j=0,k/2$ formula \newbond\
gives us $D1$-brane along the U(1) subgroup, also in accordance 
with the algebraic analysis of section 2.

\subsec{New branes on $SL(2,R)$}

\lref\bacpet{C. Bachas and M. Petropoulos, Anti-de Sitter D-branes,
JHEP 0102 (2001) 025, hep-th/0012234.}

Now let us turn to the case of $SL(2,R)$ .
A general group element can be parametrized as follows:
\eqn\grparam{g=\left(\matrix{
x_0+x_1&x_2+x_3\cr
x_2-x_3&x_0-x_1\cr}\right),}
where 
\eqn\hyp{x_0^2-x_1^2-x_2^2+x_3^2=1.}

The conjugacy class is given by the condition $x_0=C$.

Here, due to compactness of the subgroup generated by the $\sigma_2$
and non-compactness of the the subgroup generated by the 
$\sigma_3$ we have two inequivalent directions for which can be 
taken the Neumann boundary condition.

Considering first the non-compact case, if 
we parametrise the $U(1)$ subgroup as $e^{\alpha\sigma_3}$
the shape of the branes will be given by:
\eqn\branloc{e^{\alpha\sigma_3}g=\left(\matrix{
e^{\alpha}(C+x_1)&e^{\alpha}(x_2+x_3)\cr
e^{-\alpha}(x_2-x_3)&e^{-\alpha}(C-x_1)\cr}\right)=
\left(\matrix{
\tilde{x}_0+\tilde{x}_1&\tilde{x}_2+\tilde{x}_3\cr
\tilde{x}_2-\tilde{x}_3&\tilde{x}_0-\tilde{x}_1\cr}\right).}
We see that 
\eqn\condone{\tilde{x}_2^2-\tilde{x}_3^2=x_2^2-x_3^2.}

Rewriting eq.\hyp\ in the form 
\eqn\hypp{x_3^2-x_2^2=(1-C^2)+x_1^2,} 
and using \condone\
we can describe the branes by the following inequality:
\eqn\brconsl{\tilde{x}_3^2-\tilde{x}_2^2\geq 1-C^2.}
This inequality can be simplified using the Euler angle parametrisations
described in the appendix.
In the patch given by the parametrization (A.1)
it can be written as
\eqn\fiusbr{\sin^2{\theta\over 2}\geq 1-C^2,}
or 
\eqn\fiusbrc{\cos\theta\leq 2C^2-1.}
 In the patch given by formulae (A.3) it can be written as
\eqn\secusbr{-\sinh^2\tau\geq 1-C^2,}
or 
\eqn\secusbrc{\cosh\tau\leq 2C^2-1.}

Now let us turn to the case when we choose the Neumann boundary condition
 for the subgroup generated by the $\sigma_2$.
Parametrising now the $U(1)=e^{i\alpha\sigma_2}$ , we find that the branes
are located at:
\eqn\brcomploc{e^{i\alpha\sigma_2}\left(\matrix{
C+x_1&x_2+x_3\cr
x_2-x_3&C-x_1\cr}\right)=
\left(\matrix{
\tilde{x}_0+\tilde{x}_1&\tilde{x}_2+\tilde{x}_3\cr
\tilde{x}_2-\tilde{x}_3&\tilde{x}_0-\tilde{x}_1\cr}\right),}
where 
\eqn\noncompcors{\eqalign{
&\tilde{x}_0=C\cos\alpha-x_3\sin\alpha\cr
&\tilde{x}_1=x_1\cos\alpha+x_2\sin\alpha\cr 
&\tilde{x}_2=x_2\cos\alpha-x_1\sin\alpha\cr
&\tilde{x}_3=x_3\cos\alpha+C\sin\alpha.}}
We see that

\eqn\nonbrloctwo{\tilde{x}_1^2+\tilde{x}_2^2=x_1^2+x_2^2 .}

Rewriting eq. \hyp\ in the form
\eqn\twfirlb{x_1^2+x_2^2=(C^2-1)+x_3^2 ,}
and using \nonbrloctwo\ we see that this brane can be described by 
the inequality:
\eqn\ineqtwfr{\tilde{x}_1^2+\tilde{x}_2^2\geq C^2-1 .}
Using now parametrisation (A.5) we get for the brane location:
\eqn\usnon{\cosh\rho\geq 2C^2-1 .}

\newsec{Discussion}
\lref\boris{P. Bordalo, S. Ribault and C. Schweigert, 
Flux stabilization in compact groups, JHEP 0110 (2001) 036,
hep-th/0108201.}
\lref\emss{S. Elitzur, G. Moore, A. Schwimmer, N. Seiberg,
Remarks on the canonical quantization of the Chern-Simons-Witten theory,
Nucl. Phys. B326 (1989), 104.}
\lref\almal{A. Alekseev and A. Malkin, 
Symplectic structure of the moduli space of flat connections on a Riemann
surface, Commun. Math. Phys.
169 (1995), 99, hep-th/9312004.}

Here we outline some directions for the future work,
which may further clarify properties of the T-dualized branes.

1. As was noted in  \stanold\ and \boris\ if action for the boundary WZW model
is given in the form \moactbound\ , it actually fixes also the two-form
field strength  on the D-brane world-volume by the formula:
\eqn\fielstr{2\pi F=\omega^{(2)}-B,}
and the Born-Infeld action correspondingly has the form:
\eqn\dbiact{S=\int\sqrt{{\rm det}(G+\omega^{(2)})}.}
If , for example to use corresponding formulae for the case of 
the maximally symmetric conjugacy class, we will get, as found
in \bdsnew\ , brane-stabilizing magnetic monopole.
Using the formula \wzonbound\ we can compute the two-form
field strength also for the new branes. 
This observation may be can help us firmly establish
the stability properties of the branes mentioned in section 2.

2. In \egkrs\ it was noted that the conjugacy classes also arise
as intersection of the $D4$ and $D6$-branes with $SU(2)$
group manifold in the near-horizon limit of the $NS5$-brane. 
It is easy to check that corresponding intersection of the $D$-brane 
with the $NS5$-brane wrapping whole $SU(2)$, or by other words containing
 all directions
transverse to the $NS5$-brane, breaks all supersymmetries. 
In any case, it  may be possible to find
some intersection of that kind which is nevertheless stable.
This also provides indirect evidence of the stability of the whole group
 covering branes.

3. It would be interesting to construct explicitly
boundary state for the $SL(2,R)$ describing the new branes found in section 4.

\bigskip 
\noindent{\bf Acknowledgments:}
Author thanks C. Bachas, S. Elitzur and A. Giveon for useful discussions.
Author is especially grateful to S. Elitzur for a careful reading
of the manuscript and discussion in section 4.

\appendix{A}{ Euler angles for the SL(2,R)}

Let us write for further applications connection of \grparam\
to the Euler parametrisations \vil\ .
For the case of the $SL(2,R)$ one has the different Euler parametrisations 
covering different patches of the group.
One convenient Euler parametrisation of $g\in SL(2,R)$ is
\eqn\fireu{
g=e^{\chi{\sigma_3\over 2}}e^{i\theta{\sigma_2\over 2}}
e^{\phi{\sigma_3 \over 2}}.}
It is connected to \grparam\ by formulae
\eqn\firgr{\eqalign{
& x_0=\cos{\theta\over 2}\cosh{\chi+\phi\over 2}\cr
& x_1=\cos{\theta\over 2}\sinh{\chi+\phi\over 2}\cr
& x_2=\sin{\theta\over 2}\sinh{\chi-\phi\over 2}\cr
& x_3=\sin{\theta\over 2}\cosh{\chi-\phi\over 2} .}}
We see that $x_0^2-x_1^2=\cos^2{\theta\over 2}$ and 
$x_3^2-x_2^2=\sin^2{\theta\over 2}$.
The second Euler parametrisation is
\eqn\seceur{
g=e^{\chi{\sigma_3\over 2}}e^{\tau{\sigma_1\over 2}}
e^{\phi{\sigma_3 \over 2}}.}
It is connected to \grparam\ by formulae
\eqn\secgr{\eqalign{
& x_0=\cosh{\tau\over 2}\cosh{\chi+\phi\over 2}\cr
& x_1=\cosh{\tau\over 2}\sinh{\chi+\phi\over 2}\cr
& x_2=\sinh{\tau\over 2}\cosh{\chi-\phi\over 2}\cr
& x_3=\sinh{\tau\over 2}\sinh{\chi-\phi\over 2} .}}
We see that $x_0^2-x_1^2=\cosh^2{\tau\over 2}$ and 
$x_3^2-x_2^2=-\sinh^2{\tau\over 2}$.
The last Euler parametrization which we use is
\eqn\lasteur{
g=e^{i\chi{\sigma_2\over 2}}e^{\rho{\sigma_3\over 2}}
e^{i\phi{\sigma_2 \over 2}},}
which is connected to \grparam\ by formulae
\eqn\lastgr{\eqalign{
& x_0=\cosh{\rho\over 2}\cos{\chi+\phi\over 2}\cr
& x_1=\sinh{\rho\over 2}\cos{\chi-\phi\over 2}\cr
& x_2=-\sinh{\rho\over 2}\sin{\chi-\phi\over 2}\cr
& x_3=\cosh{\rho\over 2}\sin{\chi+\phi\over 2} .}}
We see that $x_0^2+x_3^2=\cosh^2{\rho\over 2}$ and 
$x_1^2+x_2^2=\sinh^2{\rho\over 2}$.
These coordinates also are known as the cylindrical coordinates.

\listrefs

\end